\documentclass[preprint,aps,amsmath,amssymb,12pt]{revtex4}
\usepackage{epsfig}
\usepackage{slashed}
\usepackage{graphicx}
\usepackage{multirow,color}
\graphicspath{{fig/}}
\begin{document}
\title{Searching for axionlike particles at future $ep$ colliders}
\author{Chong-Xing Yue}
\email{cxyue@lnnu.edu.cn}
\author{Ming-Ze Liu}
\email{liumingze19@126.com}
\author{Yu-Chen Guo}
\email{ycguo@lnnu.edu.cn}
\affiliation{
Department of Physics, Liaoning Normal University, Dalian 116029, China}

\begin{abstract}

We explore the possibility of searching for axionlike particle (ALP) at future $ep$ colliders via the subprocess $e^{-}\gamma \rightarrow e^{-}a \rightarrow e^{-}\gamma\gamma$. Sensitivities to the effective ALP-photon coupling  $g_{a\gamma\gamma}$ for its mass in the range of 10 GeV $< M_{a}< 3$ TeV are obtained for the LHeC and its high-energy upgrade, FCC-eh. Comparing to existing bounds on the ALP free parameters, we find that the bounds given by $ep$ colliders are competitive and complementary to other colliders.

\end{abstract}

\maketitle

\section*{\uppercase\expandafter{\romannumeral1}. INTRODUCTION}
Axionlike particles (ALPs) were originally motivated by the axion, which results from the dynamical solution to the strong $CP$ problem of the standard model (SM)~\cite{CP}.
ALPs are often defined as relatively light pseudoscalar particles and appear in many extensions of the SM. Both axions and ALPs are optimal candidates to explain the dark matter (DM) of the Universe~\cite{DM}.
In general, any model with global U(1) symmetry, which is spontaneously broken, predicts the existences of ALPs and their masses and couplings are independent parameters. They can couple to the SM fermions and electroweak gauge bosons via dimension-5 operators~\cite{1}. At tree level, there is no dimension-5 operator contributing to the couplings of ALP to the physical Higgs boson, which can be induced at loop level or by the high dimension operators~\cite{2}.  ALPs have anomalous couplings to gluons as optional, and they are not required to solve the strong $CP$ problem.  The experimental constraints on the effective couplings of ALP to ordinary particles have been widely studied using various experimental data from particle physics, astroparticle physics, and cosmology. Bounds obtained from the LEP and LHC in diphoton, triphoton, and monophoton final states have been summarized and partly updated in Refs.\cite{3,4,5,6,7}.

In general, the couplings of ALP to photons or $Z$ bosons can be considered independently and might be investigated separately.
The present and future collider experiments can be used to search for ALPs with masses in the broad range from eV to TeV~\cite{3,4,5,6,7,8,9,10,11,12}.
At $e^{+}e^{-}$ colliders, production  of ALP can be studied via photon fusion and ALP-strahlung in association with a photon, $Z$, or Higgs boson \cite{3}.
At hadron colliders, exotic Higgs decays and $Z$ boson decays are the most promising search channels, which have been studied in Refs. \cite{5,9}. For GeV-scale ALP produced in photon-fusion, heavy-ion collisions  at the LHC can provide strong constraints on ALP-photon couplings \cite{10}. Reference \cite{12} has studied the possibility of detecting ALP at the LHC via the process $pp\rightarrow p p\gamma\gamma$ with the subprocess $\gamma\gamma\rightarrow\gamma\gamma$. A number of these constraints are model independent and tend to vanish at high masses. It is necessary to further study the possibility of searching for ALP at upcoming or future collider experiments.

At high energies, in addition to the electromagnetic exchange, the electroweak bosons also play important roles. The $\gamma$ or $Z$ boson exchange induces neutral current deep inelastic scattering, which has been extensively explored via $ep$ collisions. In this article, we consider the possibility of searching for ALP $a$ at the LHeC \cite{LHeC} and FCC-eh \cite{FCC-eh} in a model-independent way. We assume that its mass is in the range of 10--3000 GeV and focus on the subprocess $e^{-}\gamma \rightarrow e^{-}a  \rightarrow  e^-\gamma\gamma$, in which the initial photon comes from the initial proton. The analysis of the relevant SM backgrounds and detection efficiencies of the signals are presented. Our numerical results demonstrate that, compared with other colliders, the bounds given by the LHeC on the ALP free parameters for its mass in the range of 10--100 GeV are competitive and complementary. In addition, the FCC-eh can improve the effective search limit up to 2.5 TeV.

This paper is organized as follows. After reviewing the relevant couplings of ALP to photons and $Z$ bosons, we briefly describe the theory framework in Sec. II. In Sec. III, we calculate the production cross sections of the process $e^{-}p \rightarrow e^{-}a$  at the LHeC and FCC-eh. Our analysis strategy is also discussed in this section. We finalize the prospective sensitivities of $ep$ collider experiments for the ALP parameter space before concluding in Sec. IV.

\section*{\uppercase\expandafter{\romannumeral2}. EFFECTIVE INTERACTIONS OF ALP }

The ALPs we consider are gauge singlets under the SM gauge group and are odd under $CP$. The effective interactions of ALP with the SM particles can be described by the general effective Lagrangian~\cite{1}. Among them, those that are relevant for the process $e^{-}\gamma \rightarrow e^{-}a$ appear as
\begin{eqnarray}
\mathcal{L}&=
&\frac{1}{2}(\partial_{\mu}a)(\partial^{\mu}a)-\frac{1}{2}M^{2}_{a}a^{2}-\frac{C_{BB}}{4f_{a}}aB_{\mu\nu}\tilde{B}^{\mu\nu}-\frac{C_{WW}}{4f_{a}}aW^{i}_{\mu\nu}\tilde{W}^{i,\mu\nu},
\end{eqnarray}
where $B_{\mu\nu}$ and $W^{i}_{\mu\nu}$ are the field strength tensors of the gauge groups $U(1)_{Y}$ and $SU(2)_{L}$, and we have defined the dual field strength tensors by $\tilde{B}^{\mu\nu}$ = $\frac{1}{2} \epsilon_{\mu \nu\rho\sigma}B_{\rho\sigma}$. The ALP mass $M_{a}$ and the decay constant $f_{a}$ are supposed to be free parameters. After electroweak symmetry breaking, Eq.(1) can give the couplings of ALP to the electroweak gauge bosons. The relevant terms, which are related to our calculation, are written as
\begin{eqnarray}\mathcal{L}\supset
-\frac{g_{a\gamma\gamma}}{4}aF_{\mu\nu}\tilde{F}^{\mu\nu}-\frac{g_{a\gamma Z}}{4}aF_{\mu\nu}\tilde{Z}^{\mu\nu},
\end{eqnarray}
where $F_{\mu\nu}$ and $Z_{\mu\nu}$ denote the field strength tensors of the electromagnetic field and $Z$ field, respectively, and their duals are defined as above. The couplings $g_{a\gamma\gamma}$ and $g_{a\gamma Z}$ can be written as a linear combination of the relevant free parameters
\begin{eqnarray}g_{a\gamma\gamma} = \frac{C_{BB}c^{2}_W+C_{WW}s^{2}_W}{f_{a}}, ~~~~~~~~~~~~~~~~~~~~~~g_{a\gamma Z} = \frac{2c_Ws_W(C_{WW}-C_{BB})}{f_{a}},\end{eqnarray}
where $s_{W}$=sin$\theta_{W}$ and $c_{W}$=cos$\theta_{W}$, with $\theta_{W}$ being the Weinberg angle. It is obvious that there is $g_{a\gamma\gamma}$ $\gg$ $g_{a\gamma Z}$ for $C_{WW}$ $\simeq$ $C_{BB}$. The loop-induced flavor changing processes like $B \rightarrow K a$ can give strong constraints on the coupling parameter $C_{WW}$~\cite{14}.
Thus, it is particularly interesting to consider the case $C_{WW}$ $\ll$ $C_{BB}$. From Eq.(3) we can see that there is $g_{a\gamma Z} \simeq -2~tan\theta_{W}~g_{a\gamma\gamma}$ in this case.

\section*{\uppercase\expandafter{\romannumeral3}. SEARCH FOR ALP AT $ep$ COLLIDERS}

Lepton-hadron scattering has played a crucial role in the exploration of the elementary particles over the past 60 years. After the last $ep$ collider (HERA) with the center-of-mass energy $\sqrt{s}= 318$ GeV, it is natural to consider the possibility of future $ep$ colliders. Two ideas have been discussed: the LHeC \cite{LHeC} that uses the electron beam to collide with the existing LHC beam and the FCC-eh \cite{FCC-eh} that is an option of the Future Circular Collider program. With upgrading of the LHC, the LHeC could upgrade into HE-LHeC by HE-LHC.
The electron beam collides with the 7, 13.5, and 50 TeV $p$ beams, which correspond to the LHeC, HE-LHeC, and FCC-eh, respectively.
A final LHeC run in dedicated operation could bring the total integrated luminosity close to 1 ab$^{-1}$. For the HE-LHeC and FCC-eh, we assume that the total integrated luminosity could reach 2 and 3 ab$^{-1}$, respectively.

\begin{figure}[!htb]
\begin{center}
\includegraphics [scale=0.6] {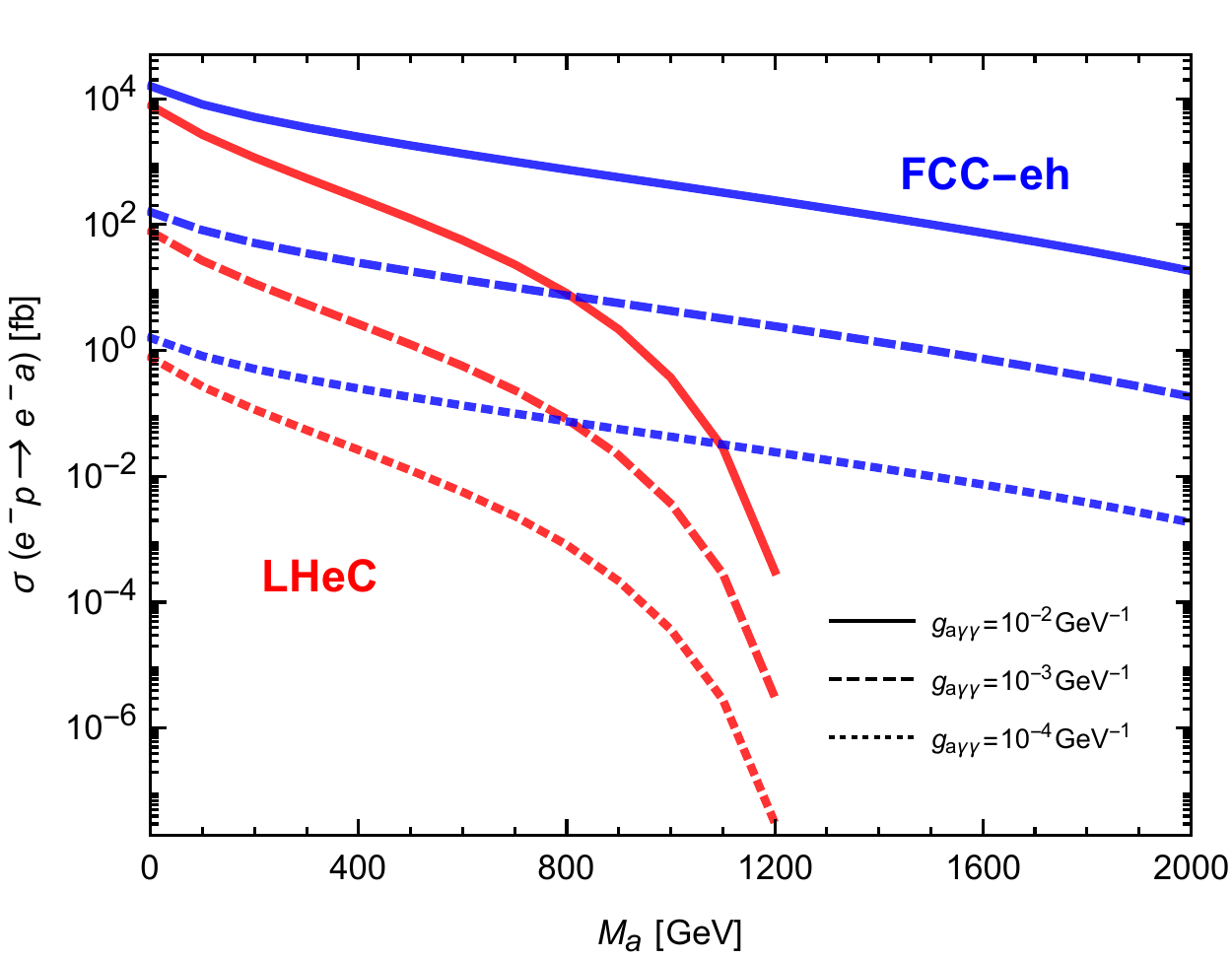}
\includegraphics [scale=0.57] {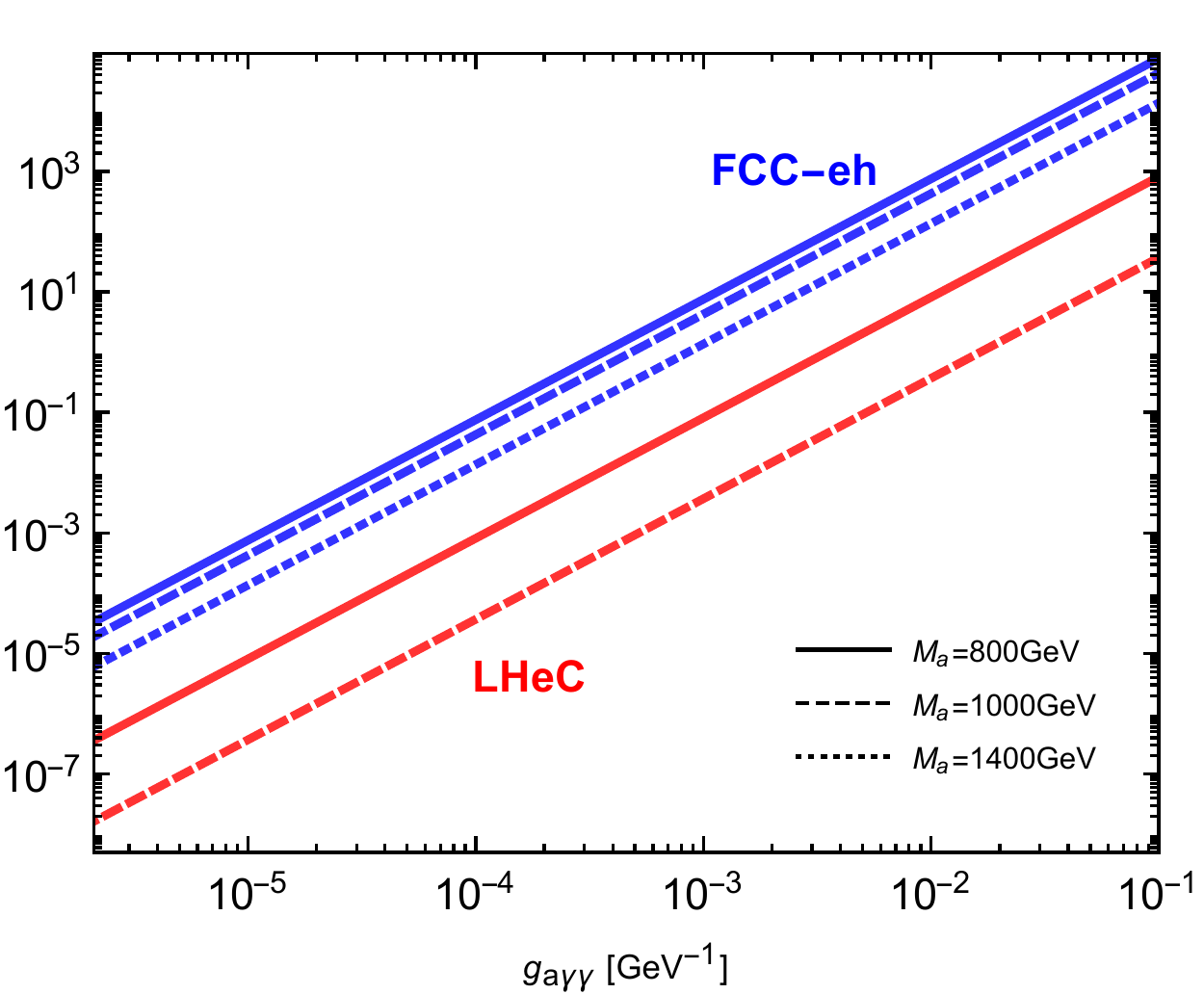}
\caption{Cross sections of the process $e^{-} p \rightarrow e^{-}a$ at the LHeC and FCC-eh as functions of $M_{a}$ \hspace*{1.4cm}(left) and $g_{a\gamma\gamma}$ (right).}\label{fig:1}
\end{center}
\end{figure}

First, we give the production cross sections for the process $e^{-}p \rightarrow e^{-}\gamma \rightarrow e^{-}a$ at the LHeC with $\sqrt{s}= 1.3$ TeV and FCC-eh with $\sqrt{s}= 3.5$ TeV. Their expressions can be uniformly written as
\begin{eqnarray}
\sigma(e^- p \rightarrow e^- a)=\int dx_1 f_{\gamma/p}(x_1,\mu^2) \cdot \hat{\sigma}(e^- \gamma  \rightarrow e^- a) \; ,
\end{eqnarray}
where the photon is emitted from the proton and can be described by the photon distribution function $f_{\gamma/p}(x, \mu^2)$.
Considering the mass range possible to be explored, we assume that the ALP mass is in the range of 10 GeV $<M_{a}<$ 1.2 TeV at the LHeC, and the mass range is broadened to 3 TeV at the FCC-eh.
The numerical results show that the production cross section for $g_{a\gamma Z}$ $\approx$ 0 is approximately equal  to that for the case of $g_{a\gamma Z}$ = -2 $tan\theta_{W}$ $g_{a\gamma\gamma}$. This is because the interference effects between the two kinds of Feynman diagrams induced by $\gamma$ and $Z$ exchanges are negative which counteract contributions of the $a\gamma Z$ coupling. Thus, we only show the cross sections in final state $e^- a$ for the case of $g_{a\gamma Z} = -2 tan\theta_{W} g_{a\gamma\gamma}$ in Fig. \ref{fig:1} as functions of the ALP mass $M_{a}$ and the coupling constant $g_{a\gamma\gamma}$.

Now we consider the possibility of searching for ALP in diphoton decay channel $e^-p \rightarrow e^{-}a \rightarrow  e^-\gamma\gamma$. The signature of the final state is characterized by the presence of a pair of photons with an invariant mass equal to the ALP mass and an isolated electron.
The final state could provide relatively high target efficiency. The SM backgrounds for this signal are dominated by the QED subprocess $e^-\gamma\rightarrow e^-\gamma\gamma$ with a real emission photon in the final state. Additional small backgrounds for small ALP mass may arise from the subprocess $e^-\gamma\rightarrow e^-\gamma$ with the third photon candidate coming from the beam-induced photon. This kind of background is reduced using the very good time resolution $\mathcal{O}$(ns) of the electromagnetic calorimeter at high photon energies \cite{LHeC}. Thus, we assume that beam backgrounds can be reduced to a negligible level without significantly affecting the signal selection efficiency.

Our event selection requires the photon with energy $E(\gamma) > 10$ GeV and  pseudorapidity $\left | \eta(\gamma) \right | < 2.5$. The invariant mass of the two photons from decays of ALP peaks close to the ALP mass. For the electron in the final state, transverse momentum $p_{T}(e) > 10$ GeV and $\left | \eta(e)  \right | < 2.5$ are required. After the basic cuts, we further employ optimized kinematical cuts according to the kinematical differences between the signal and background.
In order to carry out our numerical analysis, we use Madgraph5/aMC@NLO \cite{mg5} as the parton-level event generator, interface to the CT14QED parton distribution functions \cite{pdf}, then Pythia8 \cite{pythia} for the parton shower, hadronization. Finally, we take PGS \cite{pgs} as a detector emulator by using a LHC parameter card suitably modified for the LHeC.

\begin{figure}[!htb]
\begin{center}
\includegraphics [scale=0.35] {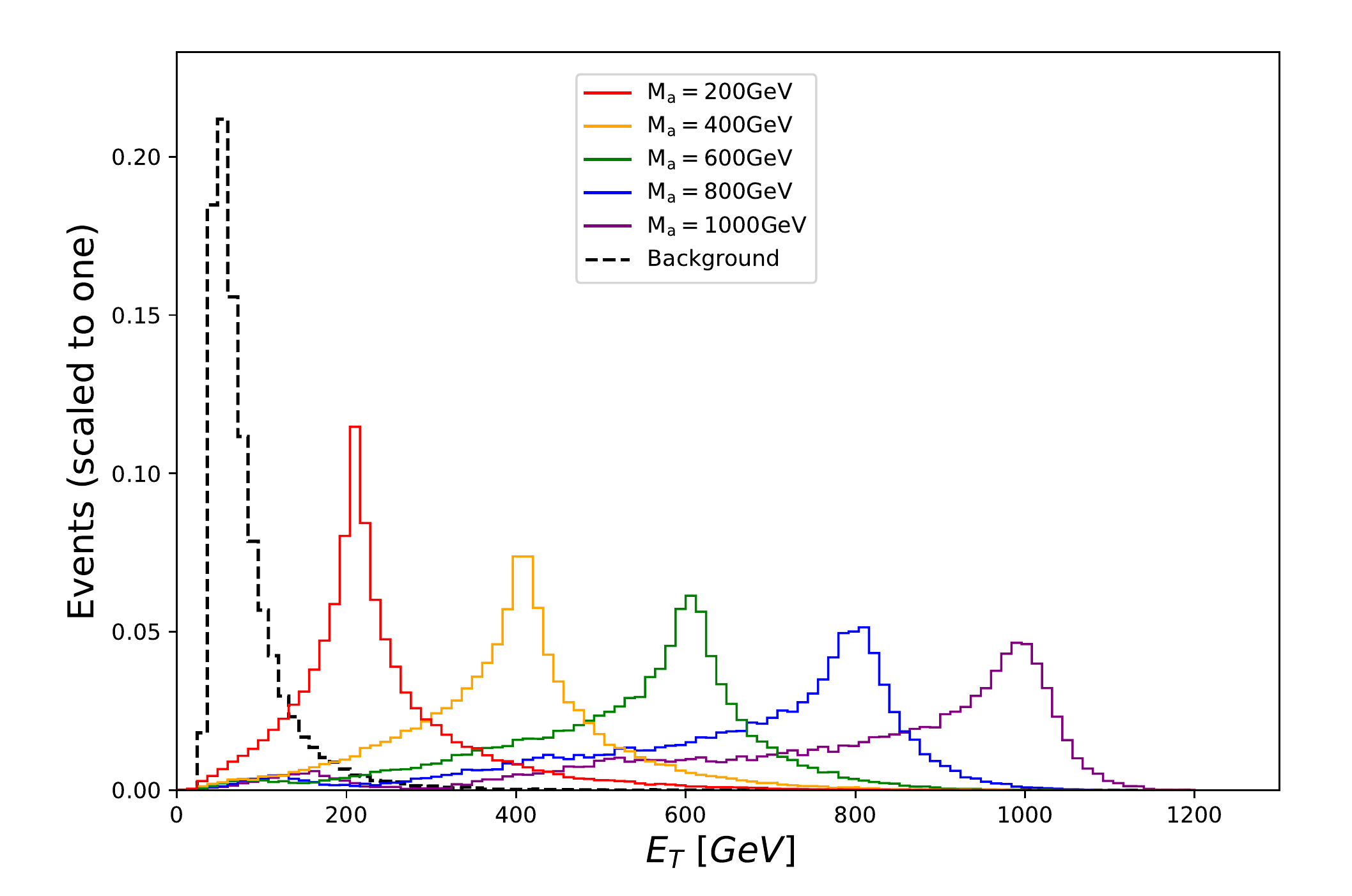}
\includegraphics [scale=0.35] {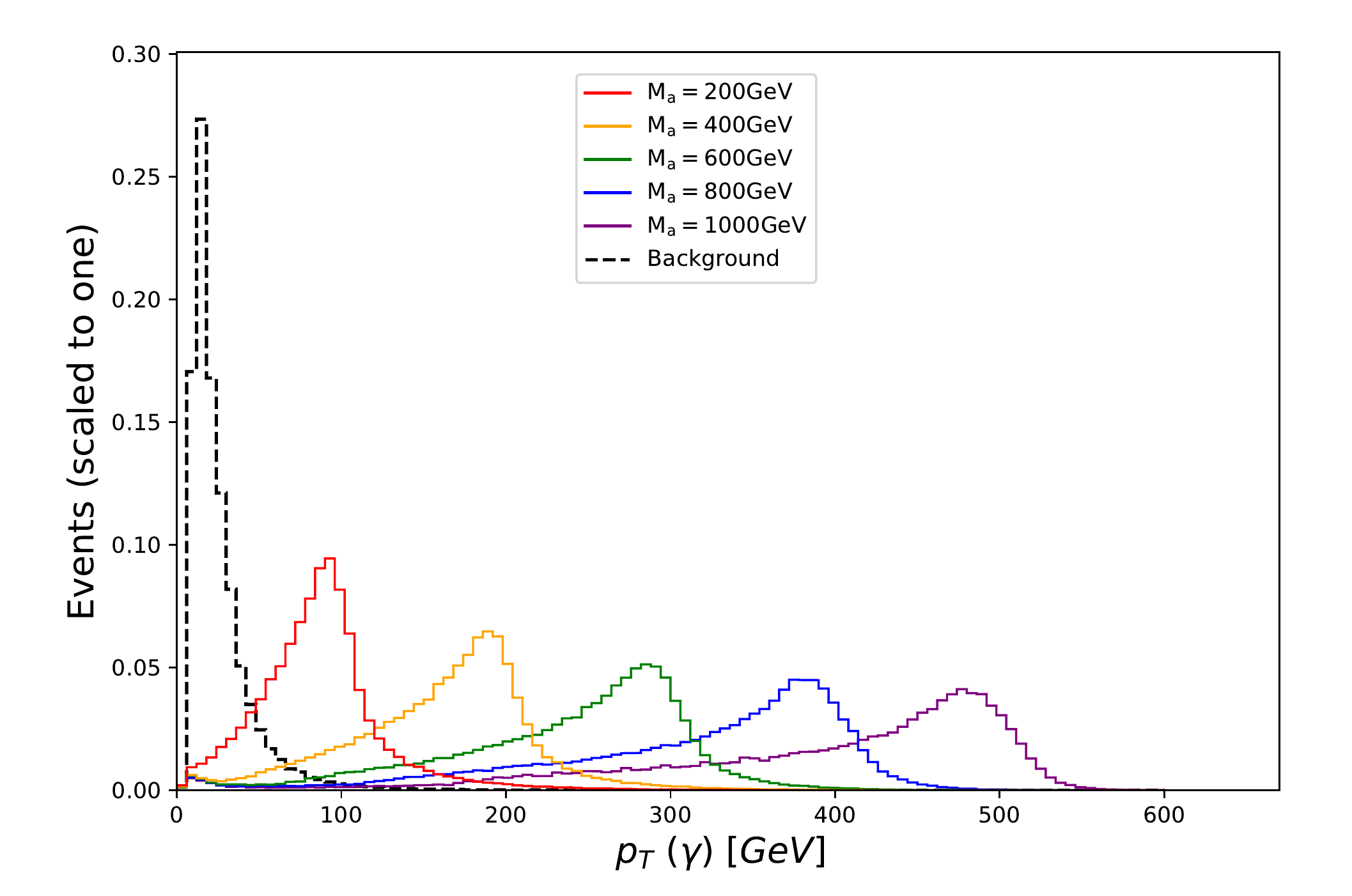}
\includegraphics [scale=0.35] {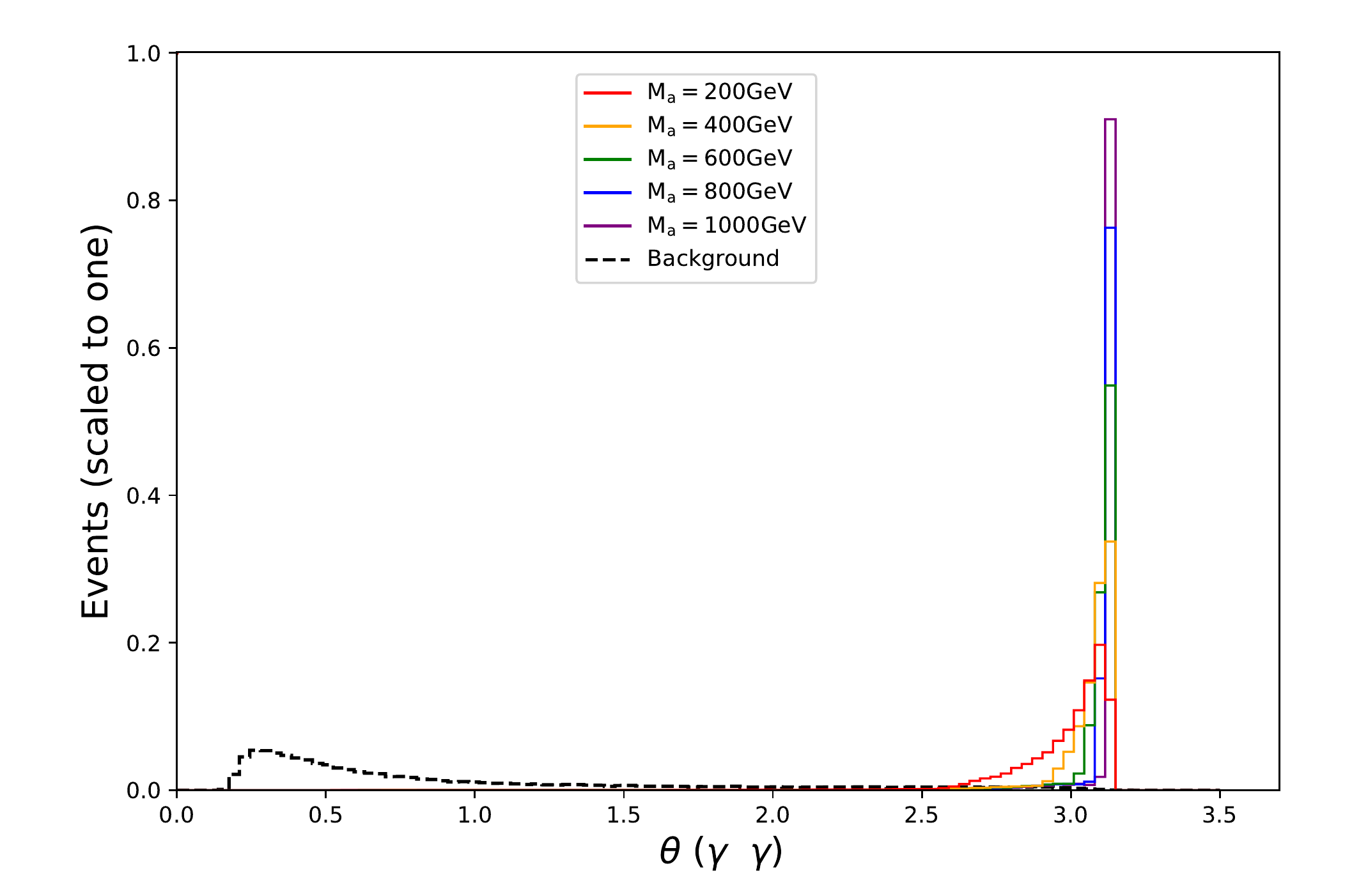}
\includegraphics [scale=0.35] {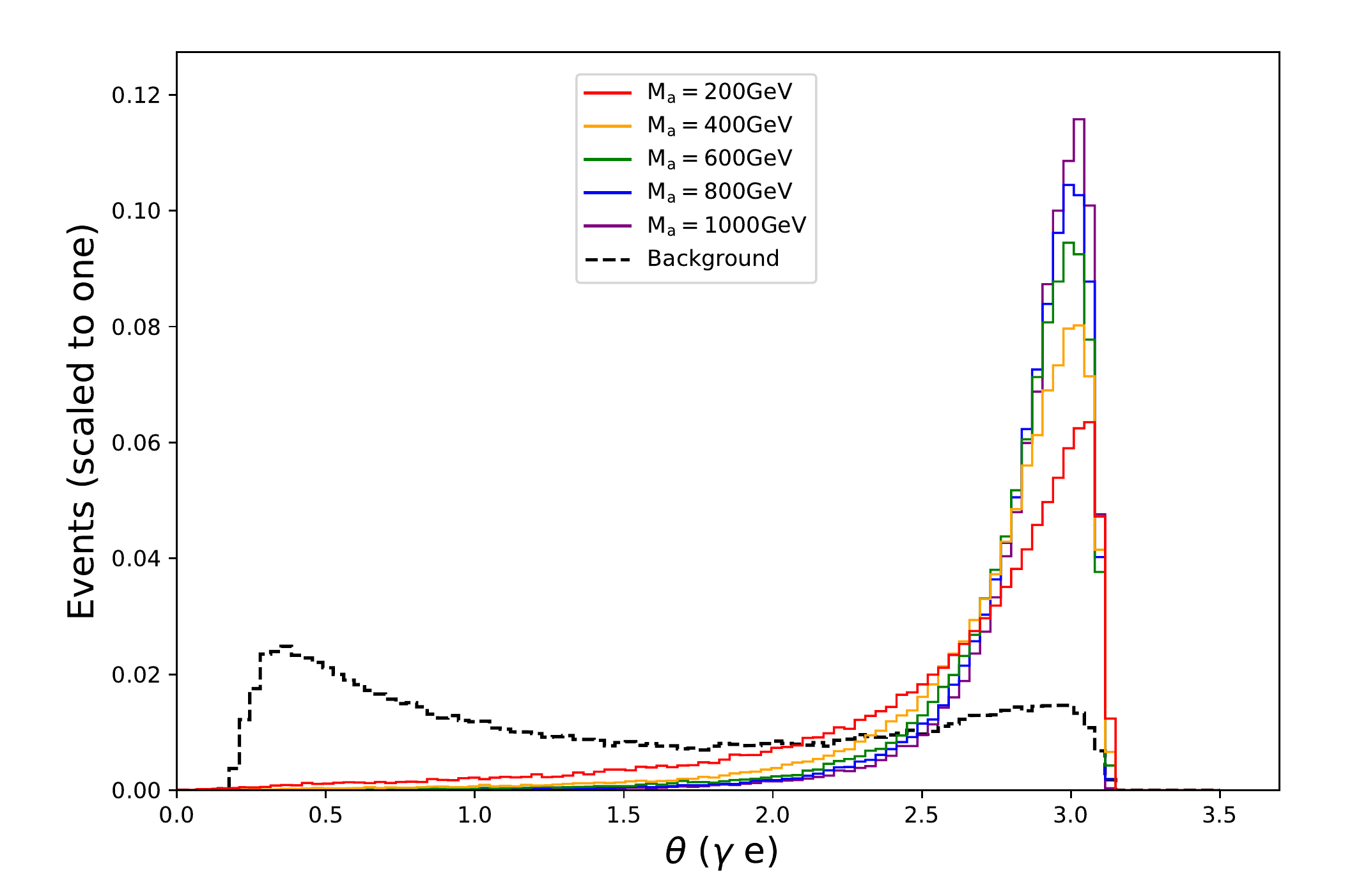}
\caption{Normalized distributions of ${E}_{T}$, $E(\gamma)$, $\theta(\gamma\gamma)$, and $\theta(\gamma e)$ from the signal and background events for different ${M}_{a}$ with $g_{a\gamma\gamma}$ $= 10^{-3}$ at  the LHeC with $\sqrt{s}= 1.3$ TeV.}\label{fig:ma5}
\end{center}
\end{figure}

For the final state of the process  $e^- p \rightarrow e^{-}a \rightarrow  e^-\gamma\gamma$, the two photons from ALP decay could be a powerful trigger. We choose to reconstruct the energy and angular distributions of the photons in the lab frame. As a result, the following kinematic variables are exploited to develop additional cuts: the angle $\theta(\gamma e)$ between the photon momentum and electron momentum, the angle $\theta(\gamma\gamma)$ between two photon momenta, and transverse momentum $p_T$ of the photon. We also apply an important global observable, the total transverse energy $E_T$. In Fig. \ref{fig:ma5}, we display the normalized distributions of these observables for some particular choices of the model parameters ($M_a=200, 400, 600, 800, 1000$ GeV with $g_{a\gamma\gamma}= 10^{-3}$) using MadAnalysis 5 \cite{ma5}. The signals are well distinguished from the corresponding backgrounds by the angle $\theta(\gamma\gamma)$. The electron momentum in the SM backgrounds is mostly along the photon direction, which is different from the signal. Just as expected, the distributions show that the $p_T(\gamma)$ spectrum peaks at around half of the ALP mass while the electrons in the SM backgrounds tend to be soft. Considering the kinematics, we impose the following improved cuts:
\begin{eqnarray}
  &p_T(\gamma) > 70 {\rm ~GeV}, ~~~\theta(e \gamma) > 2.2 ,\nonumber\\
  & E_T > 160 {\rm ~GeV}, ~~~\theta(\gamma\gamma) > 2.7.
\end{eqnarray}
These cuts could effectively remove the SM backgrounds. The event selection efficiency has been optimized with respect to the signal, and the statistical significance $\mathcal{S}=S/\sqrt{S + B}$, where $S$ and $B$ respectively denoting the numbers of signal and background events are summarized.
Here, some of the results are shown in Table \ref{table1}.

\begin{table*}[!ht]
	\caption{ Effect of individual kinematical cuts on the signals and backgrounds. The statistical significance $ \mathcal{S}$ is computed for a luminosity of 1 ab$^{-1}$, ${M}_{a} = 600$ GeV and $g_{a\gamma\gamma}= 10^{-3}$. }
	\centering
	\begin{tabular}{c|c|c|c}
		\hline
		\hline
		\multicolumn{4}{c} {LHeC,\quad $\sqrt{s}=1.3$ TeV}\\
		\hline ~~~Cuts              ~~~~~~~~~~~~~~   &~~~~~ Signal (S) ~~~~~~~~& ~~~~~ Background (B)  ~~~~~    & ~~~~~$S/\sqrt{S+B}$ \\
		\hline Initial (no cut)               &\!\!\!\!\!126     &  34910               &  0.674  \\
		\hline Basic cuts                     &116.12            &  32147.7             & 0.6465  \\
		\hline ${E}^{}_{T} > 160$ GeV    & 112.90           &   2144.3             &2.3764 \\
		\hline $\theta(\gamma e) > 2.2$     & 112.51            &   2068.5               &2.4091 \\
		\hline $p_T(\gamma) > 70$ GeV    &112.40  &  1839.1         &2.5443 \\
		\hline $\theta(\gamma~\gamma) > 2.7$   &107.77  &  443.0         &4.592 \\
		\hline
        \hline
	\end{tabular}
	\label{table1}
\end{table*}

\begin{figure}[!htb]
\begin{center}
\includegraphics [scale=0.80] {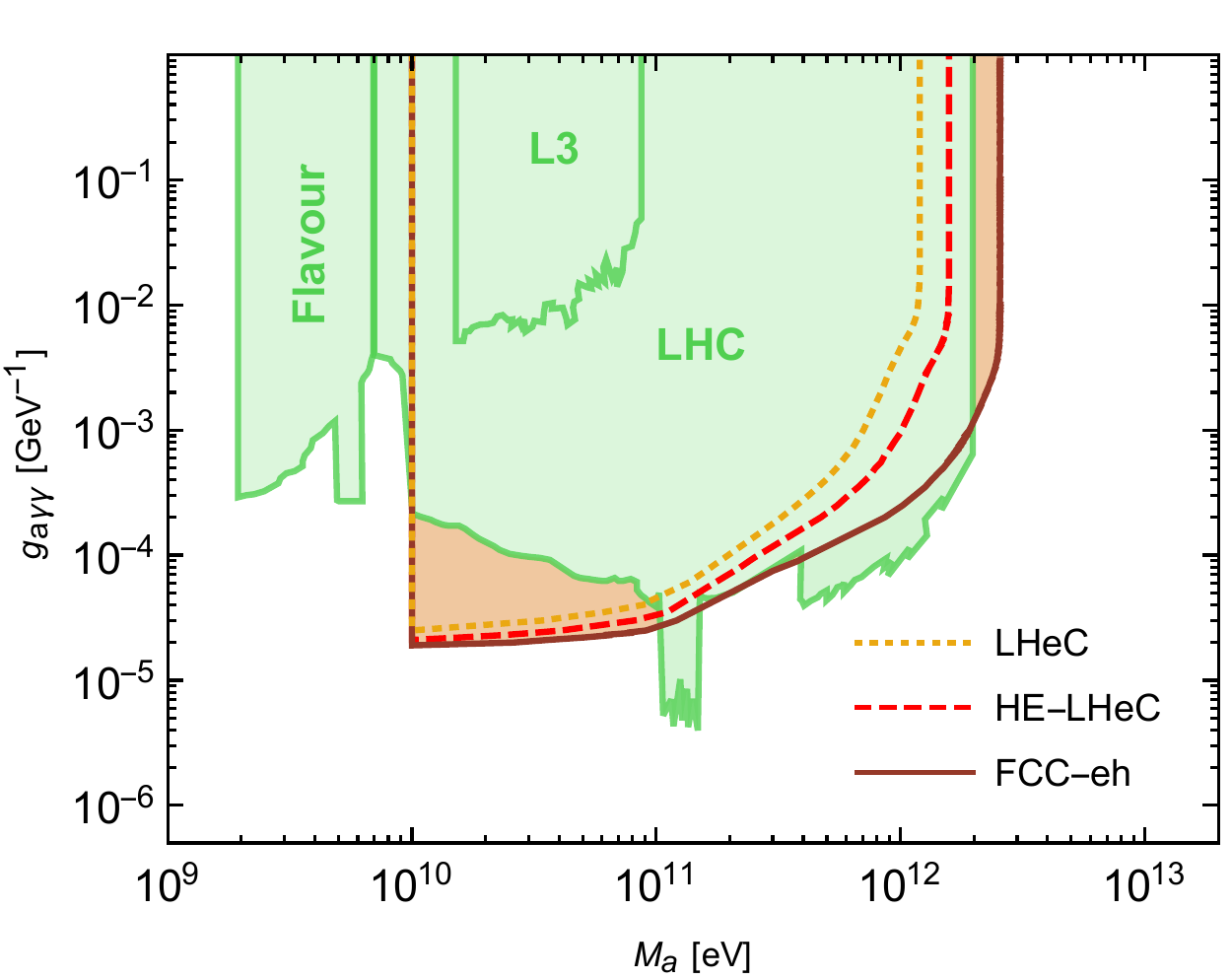}
\caption{
Projected $ep$ colliders sensitivity  at $95\%$ CL and existing constraints on the coupling of ALP with photons. The green regions are experimentally excluded.}\label{fig:K}
\end{center}
\end{figure}

Several types of experiments are used to search for ALP, ranging from the searches for direct production at colliders to those from cosmological and astroparticle physics experiments. The constraints from these searches can be mapped into the $M_{a}$-$g_{a\gamma\gamma}$ plane, and are shown in green sectors in Fig.\ref{fig:K}. The most competitive bounds for very light ALP with mass below the MeV scale come from the astrophysics and cosmology, but we consider the ALP mass range here begin with $M_a \sim 10$ GeV, for which collider experiments provide the best limits. Thus, in Fig.\ref{fig:K}, we do not show the constraints on the very light ALP. At GeV scale, Ref. \cite{7} provided the excluded parameters region by data from BABAR \cite{Babar} and LHCb \cite{LHCb}, which are adopted here and labeled flavour. For about 10 to 100 GeV, the bounds labeled L3 in Fig. \ref{fig:K} are from the analysis of Ref. \cite{1811.05466}, in which the L3 collaboration looked for hadronic final states accompanied by a hard photon \cite{L3}, though it is ultimately superseded by LHC exclusions. For the high ALP mass near the TeV scale, the limits from  data of the LHC run 1 \cite{25} are extremely strong and should be improved with the addition of run 2 data, especially at higher energies.

The projected  sensitivity contours at $95\%$ CL for the process $e^- p \rightarrow e^{-}a \rightarrow e^-\gamma\gamma$ at future $ep$ colliders are summarized in Fig. \ref{fig:K}. From this figure, one can see that, for the light ALP (i.e., $10\ \rm{GeV}< M_a < 100\ \rm{GeV}$), diphoton searches for the LHeC and FCC-eh  can push significantly beyond current constraints from existing experiments and can potentially probe the ALP-photon coupling $g_{a\gamma\gamma}$ with the order of $g_{a\gamma\gamma}\sim 10^{-5}$ to 10$^{-4}$. Furthermore, the FCC-eh will be sensitive to ALP in a large range of the parameter space and can significantly improve over existing bounds on ALP from the LHC.

\section*{\uppercase\expandafter{\romannumeral4}. CONCLUSIONS }
The existence of ALPs is a generic feature of many extensions of the SM that extend well beyond axions. Both axions and ALPs may be excellent candidates to explain the nature of DM. As pseudo-Goldstone bosons, ALPs can naturally and very weakly couple to the SM particles dominantly by couplings to photons and electroweak gauge bosons. A particular interesting decay channel  is ALP decaying into a pair of photons.

In this paper, we have investigated the search for ALP diphoton signal at future $ep$ colliders via the process $e^{-}p\rightarrow e^{-}a \rightarrow e^{-}\gamma\gamma$ in a model-independent fashion. Considering the mass range  to be possibly explored at the LHeC and FCC-eh, we focus on $10\ \rm{GeV} < M_{a} < 3$ TeV.
A proper treatment of several useful observables is presented according to the kinematical differences between the signals and relevant SM backgrounds based on the simulation performance. We apply an appropriate statistical treatment to obtain the expected bounds on the ALP free parameters. Our central observation is that existing bounds on the ALP-photon coupling for the mass interval 10--100 GeV can be significantly improved via searching for the diphoton signal at the LHeC and FCC-eh. Moreover, the FCC-eh can improve the effective search limit to 2.5 TeV. Thus, we can say that searching for ALP at future $ep$ colliders might become an important handle on new physics scenarios, which are related to ALP.

\section*{ACKNOWLEDGMENT}

\noindent
This work was supported in part by the National Natural Science Foundation of China under Grants No. 11875157, No. 11847303, and No. 11847019.

\end{document}